\documentclass[prb,a4paper,twocolumn,floatfix,showpacs,showkeys,amsmath,amssymb,nobibnotes,altaffilletter]{revtex4}

\usepackage{graphicx}% Include figure files
\usepackage{pslatex}
\usepackage{xspace}
\usepackage{subfigure}

\begin{document}

\preprint{}

\title{Origin of Reduced Polaron Recombination in Organic Semiconductor Devices}

\author{C.~Deibel}\email{deibel@physik.uni-wuerzburg.de}
\affiliation{Experimental Physics VI, Julius-Maximilians-University of W{\"u}rzburg, D-97074 W{\"u}rzburg}

\author{A.~Wagenpfahl}
\affiliation{Experimental Physics VI, Julius-Maximilians-University of W{\"u}rzburg, D-97074 W{\"u}rzburg}

\author{V.~Dyakonov}
\affiliation{Experimental Physics VI, Julius-Maximilians-University of W{\"u}rzburg, D-97074 W{\"u}rzburg}
\affiliation{Functional Materials for Energy Technology, Bavarian Centre for Applied Energy Research (ZAE Bayern), D-97074 W{\"u}rzburg}

\date{\today}

\begin{abstract}

We propose a model to explain the reduced bimolecular recombination rate found in state-of-the-art bulk heterojunction solar cells. When compared to the Langevin recombination, the experimentally observed rate is one to four orders of magnitude lower, but gets closer to the Langevin case for low temperatures. Our model considers the organic solar cell as device with carrier concentration gradients, which form due to the electrode/blend/electrode device configuration. The resulting electron concentration under working conditions of a solar cell is higher at the cathode than at the anode, and vice versa for holes. Therefore, the spatially dependent bimolecular recombination rate, proportional to the local product of electron and hole concentration, is much lower as compared to the calculation of the recombination rate based on the extracted and thus averaged charge carrier concentrations. We consider also the temperature dependence of the recombination rate, which can for the first time be described with our model.
\end{abstract}



\maketitle

\section{INTRODUCTION}

Organic bulk heterojunction solar cells have shown an increasing performance in the recent years, and also scientific progress concerning the fundamental understanding has been made.\cite{brabec2008book,park2009} However, the dominant charge carrier loss mechanism determining the photocurrent is still under discussion. The relevant processes during which the losses can occur are geminate recombination during polaron pair dissociation\cite{mihailetchi2004a,deibel2009a}, nongeminate recombination during transport of the already separated polarons\cite{shuttle2008,deibel2008b}, and charge extraction from the device.\cite{ooi2008,deibel2008a} A detailed analysis considering the interplay of these mechanisms has still to be done, as already the separate processes are  not completely described yet. Concerning nongeminate recombination, classically the bimolecular Langevin formalism\cite{langevin1903,pope1999book} has been used for low mobility materials. However, already in 1997, reports on a reduced rate as compared to  Langevin's derivation were discussed for polaron recombination in conjugated polymers,\cite{adriaenssens1997} and recently a similar reduction was found for polymer--fullerene solar cells.\cite{pivrikas2005a} Last year, we presented investigations of the polaron recombination in poly(3-hexyl thiophene):[6,6]-phenyl-C61 butyric acid methyl ester devices. We found a reduced recombination rate as compared to classic Langevin recombination, with a bimolecular decay in pristine samples, and a third order recombination in annealed samples.\cite{deibel2008b} The  low recombination rate as well as the third order decay have been observed by other researchers as well,\cite{shuttle2008,juska2008} the origin of both effects remaining unresolved. For the third order recombination, the scenario of a carrier concentration or time dependent bimolecular recombination should be considered, and is probably related to delayed recombination due to trapping in the tail of the density of states.\cite{bisquert2004} Concerning the reduced recombination rate, two models\cite{adriaenssens1997,koster2006} have been proposed in literature trying to explain the reduction mechanism. However, as we pointed out recently,\cite{deibel2008b} both fail to predict the correct temperature depenence. In this paper, we will present a simple model predicting the low bimolecular recombination rate as compared to the Langevin theory, as well as its temperature dependence.

\section{MODEL}

\subsection{Existing models for the reduced Langevin recombination}

\begin{figure}[tb]
	\includegraphics[width=7.5cm]{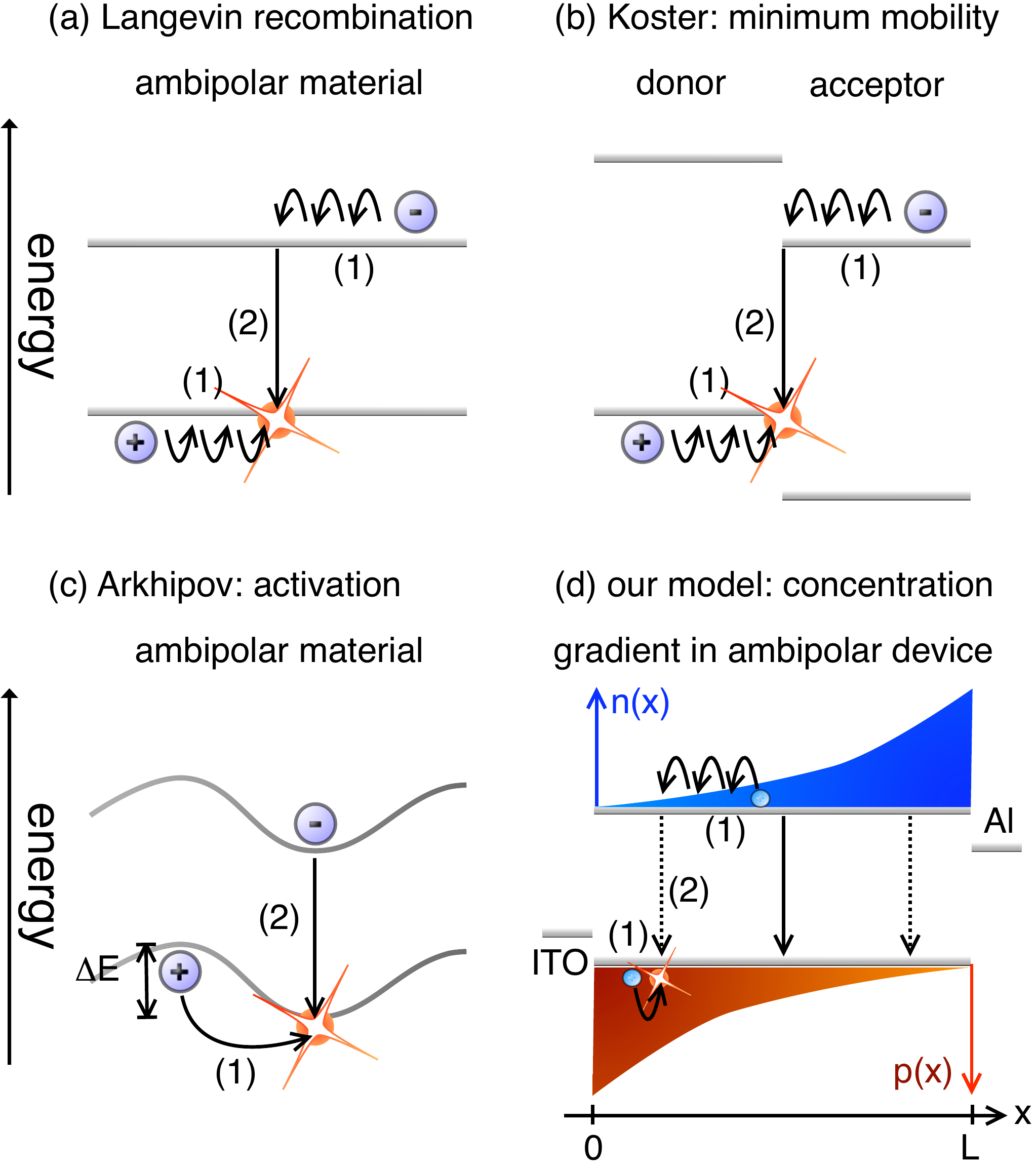}%
	\caption{(Color Online) Recombination mechanisms in low mobility materials, all being based on the Langevin recombination shown in (a). (a) to (c) consider local regions within a device, whereas (d) corresponds to the whole device of thickness $L$. $n(x)$ and $p(x)$ are the position dependent electron and hole concentration, respectively. The details are described in the text.
	\label{fig:recombination-scheme}}
\end{figure}

Before introducing our model explaining the reduced Langevin recombination, let us briefly present the basic idea behind the previously published models. Four different recombination models are shown in Fig.~\ref{fig:recombination-scheme}: (a) the classic Langevin recombination, (b) the minimum mobility model by Koster et al.\cite{koster2006}, (c) the model by Adriaenssens and Arkhipov,\cite{adriaenssens1997} and (d) our model. 

In the classic Langevin recombination (Fig.~\ref{fig:recombination-scheme}(a)), the derivation of which is nicely shown in the book by Pope and Swenberg,\cite{pope1999book} assumes that the rate limiting factor for recombination is the finding of the respective recombination partners (1), and \emph{not} the actual recombination rate (2). Neglecting process (2) as it is faster than (1), the finding of electron and hole depends on the sum of their diffusivities or---considering the Einstein relation---their mobilities. Thus, the Langevin recombination rate is
\begin{equation}
	R = \gamma (n p - n_i^2 ) ,
	\label{eqn:Langevin}
\end{equation}
where $n$ and $p$ are electron and hole concentrations, respectively, and $n_i$ is the intrinsic carrier concentration. Here, 
\begin{equation}
	\gamma=\frac{q}{\epsilon_r\epsilon_0}(\mu_e+\mu_h)
	\label{eqn:gamma}
\end{equation} 
is the Langevin recombination prefactor, where $q$ is the elementary charge, $\epsilon_r\epsilon_0$ the effective dielectric constant of the ambipolar semiconductor, $\mu_e$ and $\mu_h$ the electron and hole mobilities.

The model introduced by Koster et al.\cite{koster2006} is an extension of the Langevin model. It considers an inherent property of bulk heterojunction solar cells: the phase separation of donor and acceptor materials (Fig.~\ref{fig:recombination-scheme}(b)). Under the reasonable assumption that holes are exclusively transported in the donor polymer phase, and electrons through the fullerene acceptor, a bimolecular recombination can only take place at the heterojunction. Therefore, if the slower charge carrier does not reach the interface, no recombination takes place. In order to consider this behaviour, Koster et al.\ let the recombination prefactor be governed by the minimum mobility,
\begin{equation}
	\gamma_K=\frac{q}{\epsilon_r\epsilon_0}\text{min}(\mu_e,\mu_h) .
	\label{eqn:gamma-koster}
\end{equation} 

The Arkhipov model~\cite{adriaenssens1997} proposes that potential fluctuations in an ambipolar material are responsible for the recombination rate reduction (Fig.~\ref{fig:recombination-scheme}(c)). As the band gap remains constant, electrons and holes accumulate at the potential minima of the corresponding bands, therefore being spatially separated. In order to recombine, a potential barrier proportional to the energy difference between the minimum and maximum of the band fluctuations has to be overcome. In some respect, this model is similar to the Koster model, in as far as it also accounts for a spatial separation of the recombination partners. The origin of such a spatial separation could also be due to the above mentioned donor--acceptor phase separation. The recombination prefactor in the framework of the Arkhipov model is changed to
\begin{equation}
	\gamma_A=\frac{q}{\epsilon_r\epsilon_0}\exp\left( -\frac{\Delta E}{kT} \right) (\mu_e+\mu_h) ,
	\label{eqn:gamma-koster}
\end{equation} 
where $\Delta E$ is the activation energy, and $kT$ the thermal energy. If the Arkhipov model could be applied to bulk heterojunction solar cells, one would expect $\Delta E$ to be proportional to the energy difference between either the polymer and fullerene lowest unoccupied molecular orbitals, or between the polymer and fullerene highest occupied molecular orbitals.

\begin{figure}[tb]
	\includegraphics[width=7.5cm]{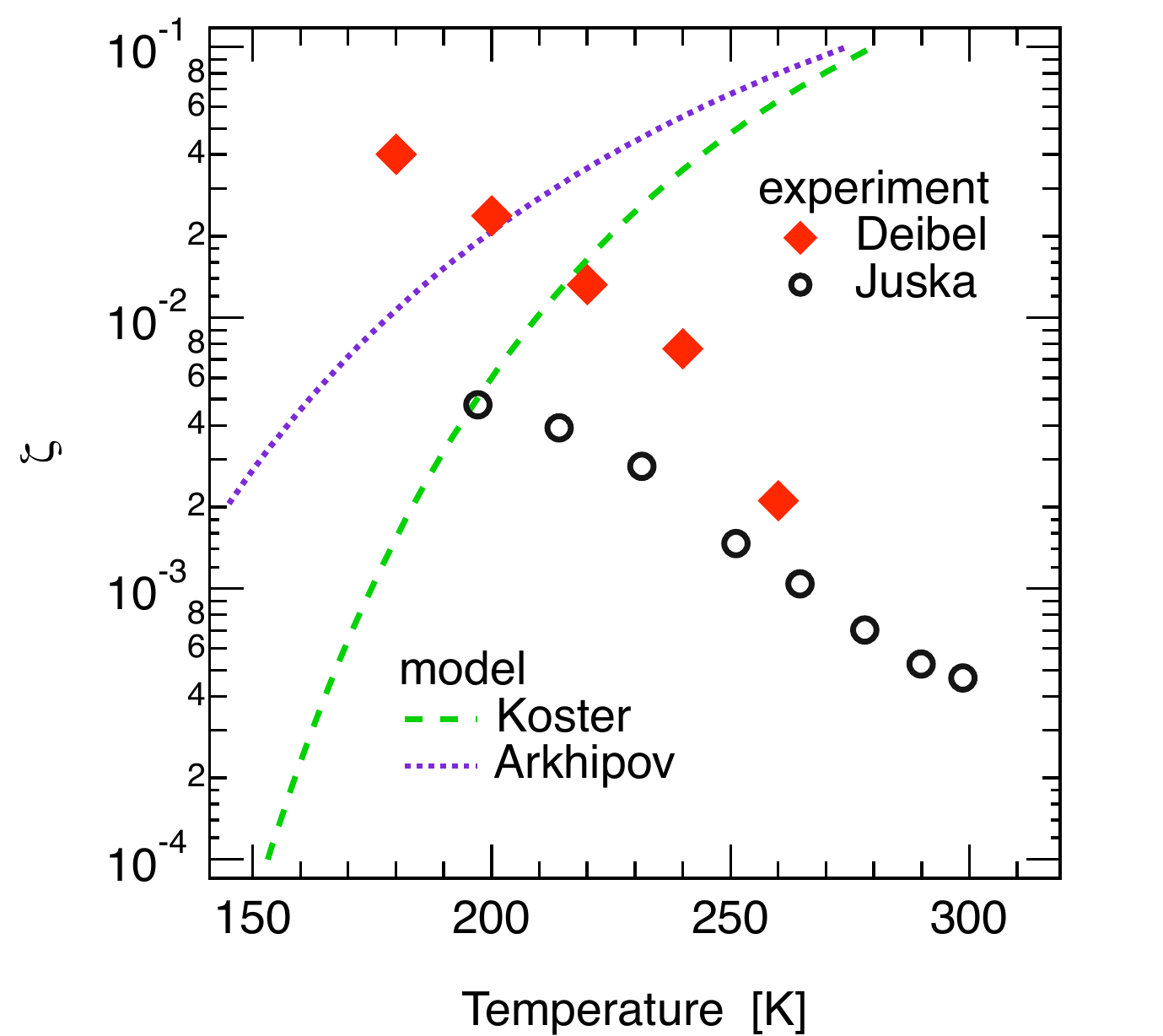}%
	\caption{(Color Online) Langevin recombination reduction factor $\zeta$ in dependence on temperature. Shown are the results for annealed devices of ours (diamonds),\cite{deibel2008b} and of Juska et al.~\cite{juska2006} (circles). The models by Koster et al.~\cite{koster2006} (dashed) and Arkhipov et al.~\cite{adriaenssens1997} (dotted) are also included; they show a markedly different temperature dependence as compared to the experimental data.
	\label{fig:zeta-exp-vs-oldmodels}}
\end{figure}

In our recent publication on bimolecular recombination,\cite{deibel2008b} we simultaneously determined the time dependent carrier concentration and charge carrier mobility in a photoinduced CELIV (charge extraction by linearly increasing voltage) experiment on pristine and annealed poly(3-hexyl thiophene):[6,6]-phenyl-C61 butyric acid methyl ester solar cells. A positive aspect of this experimental technique is that is able to determine the reduction factor
\begin{equation}
	\zeta = \frac{R}{R_\text{experiment}},
	\label{eqn:Rexp}
\end{equation}
which is the fraction of the Langevin rate $R$ (Eqn.~(\ref{eqn:Langevin})) over the experimentally determined recombination rate $R_\text{experiment}$. A similar experiment had been done previously by Juska et al.\cite{juska2006} From the experimental results shown in Fig.~\ref{fig:zeta-exp-vs-oldmodels}, $\zeta$ shows a negative temperature coefficient in annealed devices, increasing with decreasing temperature, thus minimizing the difference to the classic Langevin rate. In constrast to the experimental findings, the Arkhipov model as well as the Koster model predict a positive temperature coefficient of $\zeta$, thus increasing with temperature, going asymptotically closer to the classic Langevin rate ($\zeta=1$). As a side note, in order to calculate the temperature dependence for the Koster model, temperature dependent mobilities have to be used. A suitable model is the gaussian disorder model,\cite{bassler1993} which implies an exponentially decreasing mobility with falling temperature, depending mostly on the energetic width of the gaussian density of states, $\sigma$, also called disorder parameter. Attributing different values of $\sigma$ to the electron and hole transporting phases in a bulk heterojunction, which is in line with Koster's assumptions, the temperature dependence of Eqn.~(\ref{eqn:gamma-koster}) shows the behaviour as described above. Only in the case of having the same disorder parameter for electrons and holes does the calculated recombination prefactor $\gamma_K$ become temperature independent, but it never can attain the experimentally found temperature coefficient. Thus, neither the Juska\cite{juska2006} nor the Koster\cite{koster2006} model can predict the temperature dependence of the recombination reduction factor $\zeta$ correctly.

\subsection{Carrier concentration gradient model for the reduced Langevin recombination}

Our model considers the discrepancy between the experimental determination of the carrier concentration by charge extraction techniques, which gives only average values, and the locally varying carrier concentration gradients found in the devices under working conditions. This difference is of particular importance for bulk heterojunction devices, which consist of an ambipolar semiconductor layer---the donor--acceptor blend---sandwiched between anode and cathode. 

The experimental carrier concentration yields the complete density of charge carriers $n_\text{extracted}$ found in the device under test conditions. When calculating the Langevin recombination rate, Eqn.~(\ref{eqn:Langevin}), usually the assumption $n_\text{extracted}=n=p$ is made, so that $R=\gamma n_\text{extracted}^2$ should fit the experimental data. However, as described above, the additional reduction factor $\zeta$ had to be introduced in order to yield a good description of the experimental carrier concentration decay with time, with $R_\text{experiment} = \zeta\gamma n_\text{extracted}^2$ being the experimentally determined recombination rate.

The problem with the above mentioned assumption of $n_\text{extracted}=n=p$ stems from the implicit consequence $n(x)=p(x)$, where $x$ is the distance from anode to cathode of the device. Even if electrons and holes could be extracted separately, only spatial averages $\overline{n(x)}$ and $\overline{p(x)}$ were experimentally accessible. However, considering the charge carrier distribution in a bulk heterojunction device, these are not valid assumptions under most measurement conditions. A sketch of a typical electron and hole carrier concentration in a bulk heterojunction solar cell is shown in Fig.~\ref{fig:recombination-scheme}(d). The strong carrier concentration gradients are indeed typical for an ambipolar device with asymmetric contacts such as a bulk heterojunction solar cells. In principle, these gradients occur in the dark case and under illumination; in the latter case, the carrier concentration gradient is somewhat lower due to the photogeneration of electron--hole pairs, but nevertheless very relevant to the topic under discussion. The limiting factor for the recombination is still the finding of electron and hole (1), which is proportional to the sum of the mobilities, as described above. However, now the different electron and hole concentration gradients have to be considered. The indium tin oxide (ITO) electrode, the anode, is a good hole injection contact into conjugated polymers such as P3HT. Consequently, the hole concentration of the whole device finds its maximum at this spatial position for voltages below the built-in voltage. At the same time, hole concentration at the cathode is much lower for voltages below the flat band case, in darkness and under illumination. The concentration gradients are lowered due to illumination, as the generation of electron--hole pairs throughout the bulk changes the carrier concentration mostly where it was low without light. That means that the relative increase of the hole concentration is highest in the vicinity of the cathode, despite the extraction path of the photogenerated holes being via the anode. Thus, the considerations concerning the carrier concentration gradients apply to both, dark and illuminated devices. In order to calculate a recombination rate based on average carrier concentrations $\overline{n(x)}$ and $\overline{p(x)}$---considering these conditions---introduces a large systematic error.

In order to better illustrate the differences arising when comparing recombination rates calculated from either average carrier concentrations or actual gradients, we devised a simple model. This model will allow us to get a better impression of the recombination reduction factor $\zeta$, and thus the origin of the seemingly reduced Langevin recombination rates in ambipolar organic devices, in particular bulk heterojunction solar cells.

According to our statement, $\zeta$ can be defined as
\begin{equation}
	\zeta = \frac{\frac{1}{L} \, \int_0^L \, n(x) p(x) \,dx}{\overline{n(x)}\cdot\overline{p(x)}} ,
	\label{eqn:zeta}
\end{equation}
where the denominator corresponds to spatial averages, for instance when using carrier concentrations from charge extraction experiments. The numerator instead correctly accounts for the carrier concentration gradients found \emph{in} the device under a certain applied voltage and given light intensity.

\begin{figure}[tb]
	\includegraphics[width=7.5cm]{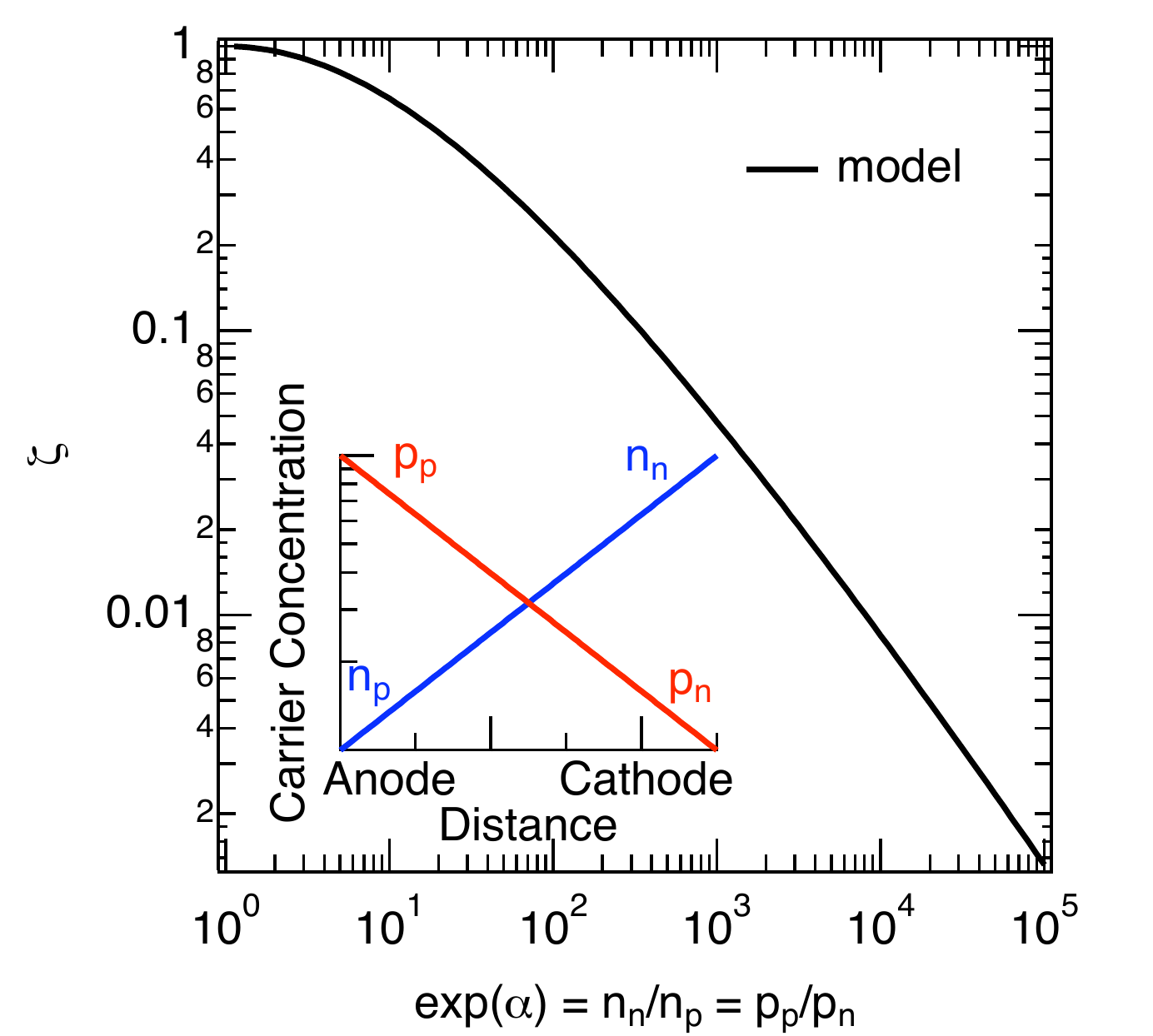}
	\caption{(Color Online) The simple model for the Langevin reduction factor $\zeta$, given by  Eqn.~(\ref{eqn:np}). It depends on the parameter $\alpha$, which represents the carrier concentration gradients. The inset shows the carrier concentration (on a logarithmic scale) vs.\ distance after Eqns.~(\ref{eqn:def-n}) and~(\ref{eqn:def-p}). $p_p$ ($n_p$)is the hole (electron) concentration at the hole injection electrode, the anode. Similarly, $n_n$ ($p_n$)is the electron (hole) concentration at the electron injection electrode, the cathode. 
	\label{fig:zeta-simple-model}}
\end{figure}

For simplicity, we define mirror-symmetric electron and hole carrier concentration gradients by 
\begin{eqnarray}
	n(x) & = & n_n \exp\left( -\alpha\frac{x}{L} \right) \label{eqn:def-n}\\
	p(x) & = & p_p \exp\left( -\alpha\frac{L-x}{L} \right) \label{eqn:def-p}.
\end{eqnarray}
Here, $\alpha = \ln (n_n/n_p) = \ln (p_p/p_n)$, where $n_n$ ($p_p$) is the electron (hole) concentration at the electron (hole) injecting contact, and $n_p$ ($p_n$) is the electron (hole) concentration at the anode (cathode). $x$ is the distance from anode ($0$) to cathode ($L$). The resulting distance dependent carrier concentrations are shown in the inset of Fig.~\ref{fig:zeta-simple-model}.

Now, we can continue the calculation started in Eqn.~(\ref{eqn:zeta}) using Eqns.~(\ref{eqn:def-n}) and~(\ref{eqn:def-p}),
\begin{eqnarray}
	\zeta & = & \frac{ \frac{1}{L}\int_0^L n(x) p(x) dx }{ \frac{1}{L}\int_0^L n(x) dx \cdot \frac{1}{L}\int_0^L p(x) dx }\\
%		& = & \frac{e^{-\alpha} n_n p_p}{\frac{\left( 1 - e^{-\alpha} \right)^2}{\alpha^2}n_n p_p} \\
		& = & \alpha^2 e^{-\alpha}\left( 1 - e^{-\alpha} \right)^{-2}
	\label{eqn:np}
\end{eqnarray}

The calculated Langevin recombination reduction factor is shown in Fig.~\ref{fig:zeta-simple-model}. The steeper the carrier concentration gradients, the larger the discrepancy to their respective spatial averages, the smaller $\zeta$. Two important consequences arise: first, the recombination reduction factor $\zeta$ depends on all parameters changing the carrier concentration gradients, such as applied voltage and charge carrier mobility. Second, $\zeta$ does not directly depend on the actual recombination mechanism: no matter if the dominant recombination is monomolecular or bimolecular, or if the solar cell is illumated or not, what counts is the resulting steady state carrier concentration.

In macroscopic device simulators considering at least one space dimension, carrier concentration gradients are already accounted for. Therefore, $\zeta$ does not need to be explicitly considered. On the contrary, as the simple model presented above uses very rudimentary functions to describe the carrier concentration gradients, not considering injection barriers etc., we will apply a macroscopic simulation program in order to better understand the apparently reduced recombination under typical measurement conditions.

\section{MACROSCOPIC SIMULATION}

\begin{table}
	\centering
		\begin{tabular}{lll}
			\hline
			\hline
			parameter & value & description \\
			\hline
			$E_{Gap}$	& $1.1~\text{eV}$ & effective
bandgap~\cite{vandewal2008,veldman2009} \\
			$\Phi_n,~\Phi_p$ & $0.1~\text{eV}$ & injection
barriers\\
			$\mu_n,~\mu_p$ &
$1\cdot10^{-8}~\text{m}^2\text{V}^{-1}\text{s}^{-1}$ & mobilities~\cite{baumann2008}\\
			$d$ & $100~\text{nm}$ & active layer thickness\\
			$G$ & $6.0\cdot10^{27}~\text{m}^{-3}\text{s}^{-1}$ &
generation rate\\
			$T$ & $300~\text{K}$ & temperature\\
			$N_{eff}$ & $ 1.0\cdot10^{26}~\text{m}^{-3} $ &
effective density of states\\		
			$\epsilon_r$ & $3.4$ & relative static
permittivity~\cite{persson2005}\\	
			\hline
			\hline
		\end{tabular}
	\caption{Parameters used in the macroscopic simulation.}
	\label{tab:param}
\end{table}

The macroscopic simulation program implemented by us solves the differential equation system of the Poisson, continuity and drift--diffusion equations by an iterative approach as described in Ref.~ \cite{deibel2008a}. Additionally, we consider injection barriers at both electrodes as well as a finite surface recombination. We use the field independent surface recombination velocity $S(0)$ of the well known Scott--Malliaras model~\cite{scott1999a} considering mirror charge effects at surfaces for both electrodes. The surface recombination current is defined as
\begin{equation}
	J_{Rec} = q S(0) \left( n - n_{th} \right)
\end{equation}
with
\begin{equation}
	S(0) = 16 \pi \epsilon \epsilon_0 \left(kT \right)^2 \mu / q^3 .
\end{equation}
Here, $n$ is the electron concentration at the surface, and $n_{th}$ is the thermally activated carrier concentration. The surface recombination current is defined for both carrier types at each electrode.

In order to clearly and unambiguously show the effect of internal charge carrier distribution and its impact on the recombination rate, we deactivated the field dependent polaron pair dissociation. Consequently, the net generation rate $U(x)$ is simply a function of the generation rate $G$ and the classical Langevin recombination as defined in Eqn.~(\ref{eqn:Langevin}),
\begin{equation}
	U\left(x\right)=G - \frac{q}{\epsilon_r\epsilon_0} \left(\mu_n +
	\mu_p\right) \left( n(x)p(x) - n_i^2 \right).
\end{equation}
The parameters assembled in Tab.~\ref{tab:param} were used for all simulations, unless explicitly mentioned. Where temperature dependent calculations were performed, we varied the mobility according to the gaussian disorder model~\cite{bassler1993} with a disorder parameter of $\sigma=75$~meV and a prefactor chosen to achieve a mobility of $10^{-8}$ m$^2$/Vs for electrons and holes at $300$K.

We point out that macroscopic simulations are very useful to study organic devices such organic bulk heterojunction solar cells, despite the assumption of an effective medium. In the latter, the hole conducting properties are derived form the donor material, whereas the electron conduction properties come from the acceptor material. For donor--acceptor blends with a very fine-grained phase separation, the assumption of an effective medium is very good. For coarser phase separations, the situation becomes more difficult, as band bending between the two phases cannot be described with the effective medium. Nevertheless, as parameters derived from microscopic Monte Carlo and Master equation simulations\cite{pasveer2005,houili2006a, groves2008a} and analytic theory~\cite{onsager1938,braun1984} can be used to describe the properties of the donor--acceptor blend, the use of macrosopic simulations offers a very good insight into the impact of microscopic charge transport and recombination properties on the macroscopic device parameters such as current--voltage characteristics. Indeed, macroscopic simulations complement the microscopic point of view very well, in particular as also asymmetries due to different work functions for electron and hole injection, and their influence on the device properties can be studied. The usefulness of this approach has been reported previously,~\cite{koster2005b,buxton2006,deibel2008a} and is in the focus of the present work as well.

\section{RESULTS AND DISCUSSION}

\begin{figure}[bt]
	\includegraphics[width=8.5cm]{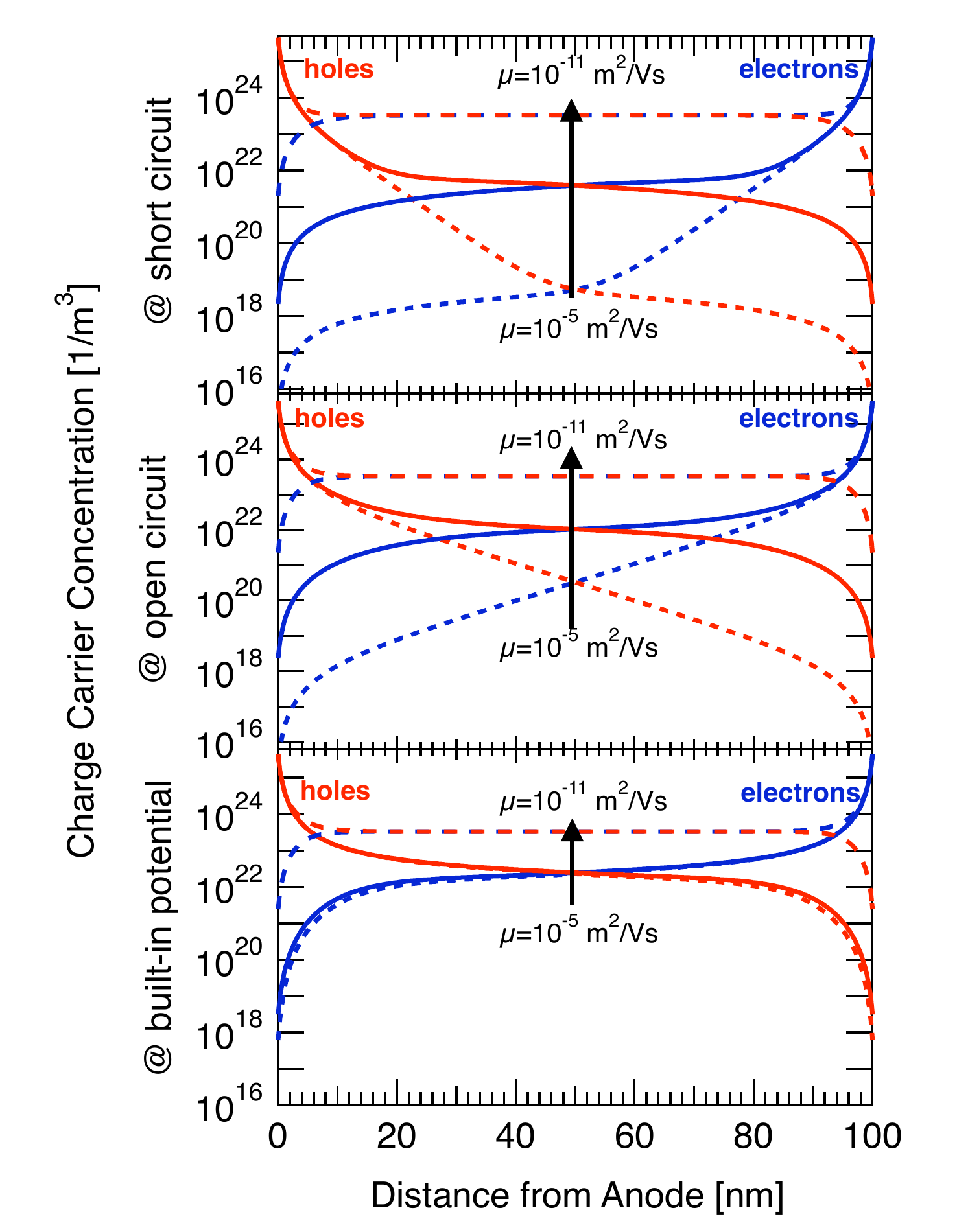}
	\caption{(Color Online) Simulated steady state electron and hole concentrations in bulk heterojunction solar cells under short circuit (top graph), open circuit (middle), and at the built-in potential (bottom) at an illumination of 1 sun. In each graph, the concentration profiles for the balanced electron--hole mobilities $\mu=10^{-5}$ (short dashed line), $10^{-8}$ (solid line), and $10^{-11}$ m$^2$/Vs (long dashed line) are shown. Holes have the highest concentration at the anode (left), electrons at the cathode (right).
	\label{fig:carrier-concentrations}}
\end{figure}

Fig.~\ref{fig:carrier-concentrations} shows the electron and hole concentrations under short circuit, open circuit, and the built-in potential. The latter is the voltage at which photo-CELIV measurements are usually performed. In comparison to our simple model (Fig.~\ref{fig:zeta-simple-model}), the concentration gradients have a more complicated shape due to injection and extraction as well as the interplay of generation and recombination in steady state. Nevertheless, it is clear that the effect remains the same. For each case shown, $\zeta$ as defined in Eqn.~(\ref{eqn:zeta}) is much smaller than unity. Consequently, charge extraction experiments such as photo-CELIV or transient photocurrents will yield the average carrier concentrations, which---if used to calculate the bimolecular recombination rates---will yield overestimated values. Looking in more detail, some important features of $\zeta$ are seen. The carrier concentration gradient is strongest at short circuit, and very low at the built in potential. In the former case, charge extraction is most favourable, whereas under flatband conditions, the charges tend to stay within the device due to the lack of a driving force. Similarly, a high mobility will tend to create steeper gradients. 

A more general feature of the bimolecular recombination in an ambipolar device is implicitly shown in Fig.~\ref{fig:carrier-concentrations}: due to the opposite electron and hole concentration profiles as well as the influence of the electrodes, the local product of electron and hole density is lowest where the deviation of the both concentrations to the average is largest---and correspondingly, $\zeta$ is smallest. This means that the polaron recombination is generally weakest at the contacts.

\begin{figure}[tb]
	\includegraphics[width=7.5cm]{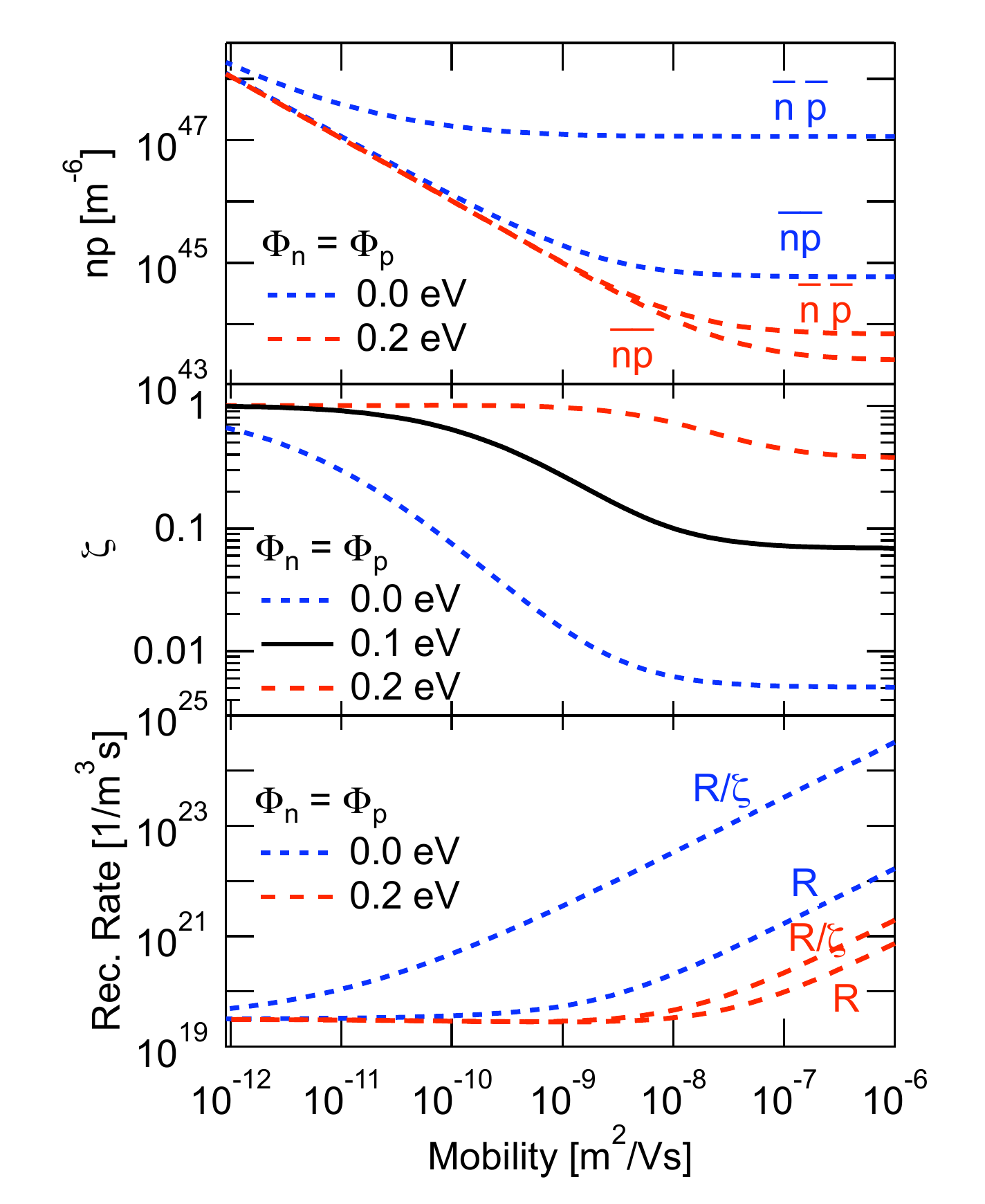}
	\caption{(Color Online) (Top) Squared carrier concentrations, by global ($\overline{np}$) or local multiplication ($\bar{n}\bar{p}$) of electron and hole concentration, respectively, in dependence on the charge carrier mobility. The data shown was calculated for flat band conditions, as they are typically used in photo-CELIV measurements. (Middle) The recombination reduction factor in dependence on the charge carrier mobility, $\zeta(\mu)$, for different injection barriers ($\Phi_p$ at the anode, $\Phi_n$ at cathode) of 0~eV (short dashed line), 0.1~eV (solid line), and 0.2~eV (long dashed line). $\zeta$ is lower for efficient extraction at high mobilities, and for low injection barriers. (Bottom) The simulated bimolecular recombination rate vs.\ the charge carrier mobility, $R(\mu)$, after Eqn.~(\ref{eqn:Langevin}), as a result of the carrier concentration gradients. Also shown is the recombination rate $R/\zeta$, which would be determined if the carrier concentration gradients were neglected.
	\label{fig:zeta-R-mu}}
\end{figure}

The discrepancy between average and local product of electron and hole concentrations in dependence on the charge carrier mobility is shown in  Fig.~\ref{fig:zeta-R-mu}(top), the resulting recombination reduction factor $\zeta$ in Fig.~\ref{fig:zeta-R-mu}(middle). Both, mobilities and injection barriers, were chosen to be symmetric for electrons and holes, but qualitatively the results hold true for asymmetric conditions as well. A high mobility corresponds to an efficient charge extraction, leading to steeper electron and hole concentration gradients, and thus a lower $\zeta$.  Similarly, the lower the injection barrier, the weaker the concentration gradients, the closer is $\zeta$ to unity. The corresponding bimolecular recombination rates are shown in Fig.~\ref{fig:zeta-R-mu}(bottom). Also included is $R/\zeta$ (Eqn.~(\ref{eqn:Rexp})), the recombination rate as derived when only considering average carrier concentrations. It equals $\gamma \overline{n}^2$, thus implying a severe overestimate of the loss rate. 

Using our macroscopic simulation, we find two other influences as well. Raising the external voltage from zero to the built-in voltage, $\zeta$ will gradually converge to unity, as the carrier concentration gradients become level when coming closer to flat band conditions. The effect of photon absorption is similar: electron--hole pair generation throughout the extent of the device leads to weaker carrier concentration gradients; consequently, the recombination reduction factor $\zeta$ approaches one for high illumination densities (not shown).

\begin{figure}[tb]
	\includegraphics[width=7.5cm]{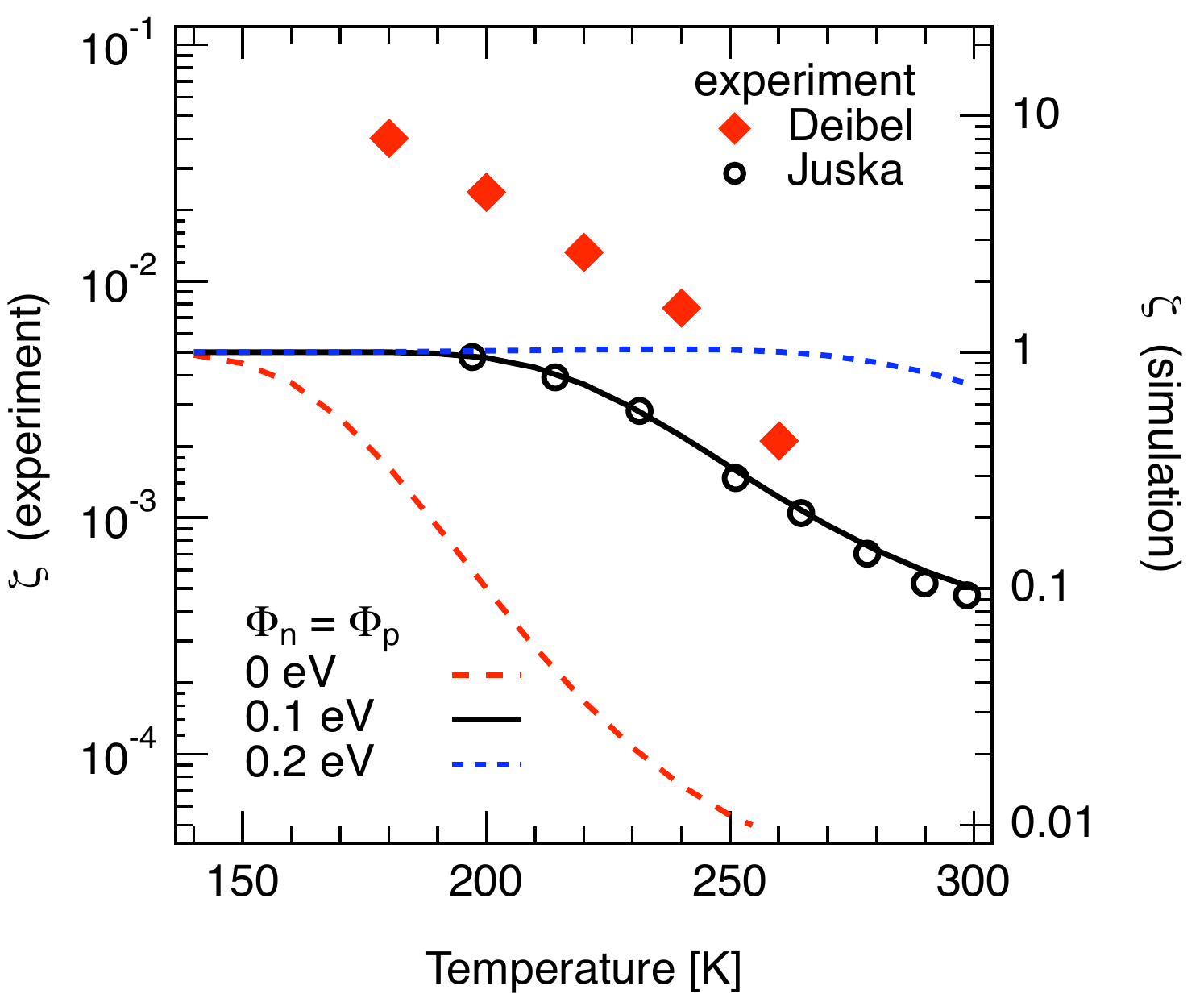}
	\caption{(Color Online) Comparison of the temperature dependent recombination reduction factor $\zeta$ for the experimental photo-CELIV data (as already shown in Fig.~\ref{fig:zeta-exp-vs-oldmodels}) with our macroscopic simulation. The difference of the absolute values of $\zeta$ in simulation and experiment, a factor of $1/200$ for the data of Juska et al.\ (circles) compares do the calculated values with an injection barrier of $0.1$~eV (black solid line). The data of Deibel et al.\ (diamonds) has similar shape, with an additional, static reduction factor of about $1/20$ is needed to match the simulation. The details are described in the text.
	\label{fig:zeta-exp-vs-sim}}
\end{figure}

%\begin{figure}[tb]
%	\includegraphics[width=7.5cm]{zeta-exp-vs-sim-v}
%	\caption{(Color Online) \xxx 
%	\label{fig:zeta-exp-vs-sim-v}}
%\end{figure}

Photo-CELIV is an experimental technique which is able to determine the correct recombination rate present in the device, despite the fact that it only considers averaged recombination rates. This is possible due to the direct fitting of the experimental time dependent carrier concentration data to the charge carrier continuity equation, 
\begin{equation}
	\frac{dn}{dt} = G(t) - \zeta R , 
\end{equation}
with the charge carrier generation rate $G(t)$, which is zero for time $t>0$ due to using a nanosecond laser pulse, the recombination reduction factor $\zeta$, and the Langevin recombination rate $R$ (Eqn.~(\ref{eqn:Langevin}). Monomolecular contributions are neglected. As photo-CELIV yields $\bar{n}$ and $\mu$ simultaneously, $R$ is completely known, and $\zeta$ can be determined.

For a comparison of our macroscopic simulation to experimental photo-CELIV data, see Fig.~\ref{fig:zeta-exp-vs-sim}. Both data and simulation are for flat band conditions, i.e., the built-in potential. 
The shape of the experimental data of Refs.~\cite{juska2006,deibel2008b} is very well reproduced by the simulations. As pointed out above, the models by Koster et al.~\cite{koster2006} and Adriaenssens and Arkhipov~\cite{adriaenssens1997} were not able to describe this temperature dependence. Thus, our model is the first one to describe the experimentally found temperature dependence of the reduced Langevin recombination qualitatively. Additionally, the voltage dependent $\zeta$ as determined by Juska et al.\cite{juska2006} corresponds to our simulations (not shown). We point out that the absolute magnitudes of simulated and experimentally determined reduction factors differs by a factor around $0.1$ to $0.005$, the discrepancy being independent of electric field and temperature. Thus, the recombination reduction factor $\zeta$ is composed of two contributions,
\begin{equation}	
	\zeta(T, F, G) = \zeta_\text{gradient}(T,F,G) \cdot \zeta_\text{static} .
\end{equation}
The first term is the temperature $T$, electric field $F$, and charge carrier generation rate $G$ dependent prefactor, which is due to the carrier concentration profiles in the device, as described by our model. The second contribution to $\zeta$ is constant, and not considered in our simulation. This static contribution $\zeta_\text{static}$ can be due one or more of the following factors: (a) a geometrical factor due to the donor--acceptor phase separation, the charges being confined to their respective phase, (b) the donor resp.\ acceptor material of the polymer--fullerene blend can have different dielectric constants,\cite{szmytkowski2009} or (c) deviations from the Langevin recombination factor due to energetic disorder, size of the donor--acceptor domains, and mismatch between the electron and hole mobility.\cite{groves2008a}

Szmytkowski~\cite{szmytkowski2009} calculates a temperature independent recombination reduction factor,
\begin{equation}
	\zeta_\epsilon = \left| \frac{\epsilon_\text{d}-\epsilon_\text{a}}{\epsilon_\text{d}+\epsilon_\text{a}} \right|
\end{equation}
with $\epsilon_\text{d}$ being the relative permittivity of the donor, and $\epsilon_\text{a}$ the corresponding value for the acceptor material. For P3HT and PCBM, the relative permittivities are approximately 3.4 and 4.0, respectively, thus $\zeta_\epsilon=0.08$, and even smaller if the respective permittivities are closer to one another. However, to our knowledge, this explanation has not been experimentally verified as of yet. 

Groves et al.~\cite{groves2008a} perform Monte Carlo simulations of electrons and holes in a blend system with hopping transport, and study deviations from the Langevin theory of bimolecular recombination inin view of  bulk devices and field effect transistors. For bulk heterojunctions, they point out that the effect of energetic disorder, domain sizes and electron--hole mobility mismatch leads to $\zeta\text{static}$ of only between $0.1$ and $1$. Therefore, the authors suggest to consider the influence of deep carrier trapping for explaining smaller recombination reduction factors. 

Thus, our model considering the carrier concentration gradients of electrons and holes in an ambipolar organic device can explain the behaviour of the reduced Langevin recombination in terms of temperature and electric field dependence, an additional static contribution is needed to match the experimental recombination rates. 

\section{CONCLUSIONS}

In conclusion, we have presented a simple model describing the reduced Langevin recombination in organic solar cells. The origin of the reduction factor is based on two contributions: first, the discrepancy between average electron and hole concentrations considered in charge extraction experiment and the usually steep carrier concentration profiles in organic semiconductors. Second, a static factor related to the phase separation of donor and acceptor, their energetic disorder and relative permittivities, the origin of which is still under discussion. Concerning the first contribution, the spatially dependent bimolecular recombination rate, proportional to the local product of electron and hole density, is much lower as compared to the rate based on average charge carrier concentrations. The latter leads to an overestimation of the recombination rates. Our model for the first time correctly describes the qualitative temperature dependence of the reduction factor found experimentally by the photo-CELIV method in P3HT:PCBM solar cells. It is also applicable to other organic electronic devices. Based on our model, and applying a device simulator, we are able to predict the voltage and light intensity dependence of the recombination reduction factor.

\begin{acknowledgments}

The current work is supported by the Bundesministerium f{\"u}r Bildung und Forschung in the framework of the OPV Stability project (Contract No.~ 03SF0334F). V.D.'s work at the ZAE Bayern is financed by the Bavarian Ministry of Economic Affairs, Infrastructure, Transport and Technology.

\end{acknowledgments}

\end{document}